\documentclass[manuscript]{acmart}

\AtBeginDocument{%
  \providecommand\BibTeX{{%
    \normalfont B\kern-0.5em{\scshape i\kern-0.25em b}\kern-0.8em\TeX}}}

\setcopyright{acmcopyright}
\copyrightyear{2021}
\acmYear{2021}
\acmDOI{10.5281/zenodo.5558085}

\acmJournal{POMACS}
\acmVolume{37}
\acmNumber{4}
\acmArticle{111}
\acmMonth{10}

\usepackage{graphicx}
\usepackage{listings}
\usepackage{subcaption}
\usepackage{booktabs}
\usepackage{multirow}
\usepackage{natbib}
\usepackage{mdframed}
\usepackage{listings}
\usepackage{soul}
\usepackage{xcolor}

\lstdefinelanguage{JavaScript}{
	keywords={typeof, new, true, false, catch, function, return, null, catch, switch, var, const, let, async, await, if, in, while, do, else, case, break, from},
	ndkeywords={class, export, boolean, throw, implements, import, this},
	sensitive=false,
	comment=[l]{//},
	morecomment=[s]{/*}{*/},
	morestring=[b]',
	morestring=[b]"
}

\usepackage{xcolor}
\definecolor{diffstart}{named}{lightgray}
\definecolor{diffincl}{named}{blue}
\definecolor{diffrem}{named}{red}

\lstdefinelanguage{diff}{
    basicstyle=\scriptsize\ttfamily,
	morecomment=[f][\color{diffstart}]{@@},
	morecomment=[f][\color{diffincl}]{+\ },
	morecomment=[f][\color{diffrem}]{-\ },
	morestring=[b]',
	morestring=[b]",
	keywords={typeof, new, true, false, catch, function, return, null, catch, switch, var, const, let, async, await, if, in, while, do, else, case, break, from},		
	ndkeywords={class, export, boolean, throw, implements, import, this},
}

\begin{document}

\title{I depended on you and you broke me: An empirical study of manifesting breaking changes in client packages}


\author{Daniel Venturini}
\email{danielventurini@alunos.utfpr.edu.br}
\affiliation{%
  \institution{Federal University of Technology (UTFPR)}
  \city{Campo Mourão}
  \state{Paraná}
  \country{Brazil}
}

\author{Filipe Roseiro Cogo}
\email{filipe.cogo@gmail.com}
\affiliation{%
  \institution{Huawei Technologies}
  \city{Kingston}
  \country{Canada}}

\author{Ivanilton Polato}
\email{ipolato@utfpr.edu.br}
\affiliation{%
 \institution{Federal University of Technology (UTFPR)}
  \city{Campo Mourão}
  \state{Paraná}
  \country{Brazil}}

\author{Marco A Gerosa}
\email{Marco.Gerosa@nau.edu}
\affiliation{%
  \institution{Northern Arizona University (NAU)}
  \state{Arizona}
  \city{Flagstaff}
  \country{United States}}
  
\author{Igor Scaliante Wiese}
\email{igor@utfpr.edu.br}
\affiliation{%
 \institution{Federal University of Technology (UTFPR)}
  \city{Campo Mourão}
  \state{Paraná}
  \country{Brazil}}
\renewcommand{\shortauthors}{Venturini, et al.}


\begin{abstract}
Complex software systems have a network of dependencies. Developers often configure package managers (e.g., \textsf{npm}) to automatically update dependencies with each publication of new releases containing bug fixes and new features. When a dependency release introduces backward-incompatible changes, commonly known as \textit{breaking changes}, dependent packages may not build anymore. This may indirectly impact downstream packages, but the impact of breaking changes and how dependent packages recover from these breaking changes remain unclear. To close this gap, we investigated the manifestation of breaking changes in the \textsf{npm} ecosystem, focusing on cases where packages' builds are impacted by breaking changes from their dependencies. We measured the extent to which breaking changes affect dependent packages. Our analyses show that around 12\% of the dependent packages and 14\% of their releases were impacted by a breaking change during updates of non-major releases of their dependencies. We observed that, from all of the manifesting breaking changes, 44\% were introduced both in minor and patch releases, which in principle should be backward compatible. Clients recovered themselves from these breaking changes in half of the cases, most frequently by upgrading or downgrading the provider's version without changing the versioning configuration in the package manager. We expect that these results help developers understand the potential impact of such changes and recover from them.
\end{abstract}


\maketitle


\section{Introduction}
\label{intro}

Complex software systems are commonly built upon dependency relationships in which a \emph{client package} reuses the functionalities of \emph{provider packages}, which in turn depend on other packages. To automate the process of installing, upgrading, configuring, and removing dependencies, package managers such as \textsf{npm}, \textsf{Maven}, \textsf{pip}, and \textsf{Cargo} are widely adopted. Despite the many benefits brought by the reuse of provider packages, one of the main risks client packages face is \emph{breaking changes}~\cite{bc_maven}. Breaking changes are backward-incompatible changes performed by the provider package that renders the client package build defective (e.g., a change in a provider's API). When client packages configure package managers to automatically accept updates on a range of provider package versions, the breaking change will have the serious consequence of catching clients off guard. For example, in \textsf{npm}, where most of the packages follow the Semantic Versioning specification~\cite{techincal_lag_of_depencencies}, clients adopt configurations that automatically update minor and patch releases of their providers. In principle, these release types should not contain any breaking changes, as the semantic version posits that only major updates should contain breaking changes. However, minor or patch releases occasionally introduce breaking changes and generate unexpected errors in the client packages when these breaking changes manifest on clients. Due to the transitive nature of the dependencies in package managers, unexpected breaking changes can potentially impact a large proportion of the dependency network, preventing several packages from performing a successful build.

Research has shown that providers occasionally incorrectly use the Semantic Versioning specification~\cite{noregrets2018}. In the \textsf{npm} ecosystem, prior research has shown that provider packages indeed publish releases containing breaking changes~\cite{model_based_bc_nodejs, using_others_tests, noregrets2018, detecting_bc_JavaScript_apis}. However, such studies provide limited information regarding the prevalence of these breaking changes, focusing on API breaking changes without clarifying how the client packages solve the problems they cause. In this paper, we fill this gap by conducting an empirical study of \textsf{npm} projects hosted on \textsf{GitHub}, verifying the frequency and types of the breaking changes that manifest as defects in client packages and how clients recover from them. \textsf{npm} is the main package manager for the \textsf{JavaScript} programming language, with more than one million packages. An estimated 97\% of web applications come from \textsf{npm}~\cite{npm_blog_2018}, making it the most extensive dependency network~\cite{package_dependency_network}. We employed mixed methods to identify and analyze the types of \emph{manifesting breaking changes} -- changes in a provider release that render the client's build defective -- and how client packages deal with them in their projects. This paper does not study cases in which a breaking change does not manifest itself in other projects. Our research answers the following questions:

\noindent
\textbf{RQ1. To what extent do breaking changes manifest themselves in client packages?}

\noindent \textit{We analyzed 384 packages selected using a random sampling approach (95\% confidence level and ±5\% confidence interval) to select client packages with at least one provider}. We found that manifesting breaking changes impacted 11.7\% of all client packages (regardless of their releases) and 13.9\% of their releases. In addition, 2.6\% of providers introduced manifesting breaking changes.

\noindent \textbf{RQ2. What changes in the provider packages manifest a breaking change?} 

\noindent \textit{The main causes of manifesting breaking changes were feature modifications, change propagation among dependencies, and data type modifications. We also verified that an equal proportion of manifesting breaking changes was introduced in minor and patch releases (approximately 44\% in each release type). Providers fixed most of the manifesting breaking change cases introduced in minor and patch releases (46.4\% and 61.5\%, respectively). Finally, manifesting breaking changes were documented in issue reports, pull requests, or changelogs in 78.1\% of cases.}

\noindent
\textbf{RQ3. How do client packages recover from manifesting breaking changes?} 

\noindent \textit{Client packages recovered from manifesting breaking changes in 39.1\% of the cases, and their recovery took about 134 days when providers did not fix the break or when clients recovered first. When providers released a fix to a manifesting breaking change, they took a median of seven days. Upgrading the provider is the most frequent way client packages recover from a manifesting breaking change.}

This paper contributes to the literature by providing quantitative and qualitative empirical evidence about the phenomenon of manifesting breaking changes in the \textsf{npm} ecosystem. Our qualitative study may help developers understand the types of changes that manifest defects in client packages and which strategies are used to recover from breaking changes. We also provide several suggestions about how clients and providers can enhance the quality of their release processes. As an additional contribution, we created pull requests for real manifesting breaking change cases that had not yet been resolved, half of which were merged.


\section{Definitions, Scope and Motivating Examples}
\label{sec:motivating}

This section defines terms used in this paper as well as describes motivating examples for our research.

\subsection{Glossary definitions}
\label{subsec:definitions}
In the following, we describe the terms and definitions that we use in the paper, based on related work~\cite{downgrades_by_filipe,npm-seven,tapir2021}.

\begin{figure}[H]
	\centering
	\includegraphics[scale=0.65]{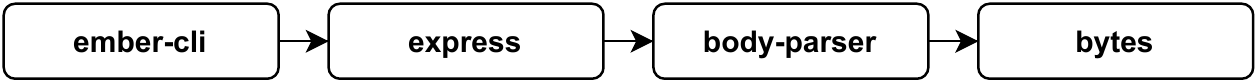}
	\caption{An example of a dependency tree, with clients and providers.}
	\label{fig:definitions_example}
\end{figure}{}

\begin{itemize}
    \item \textbf{provider package release} is the package release that provides features and resources for use by others packages releases. In Figure \ref{fig:definitions_example}, the package \textsf{express} is a provider of \textsf{ember-cli}, \textsf{body-parser} is a provider of \textsf{express}, and so on. We refer to a provider package $P$ as a transitive provider when we want to emphasize that $P$ has other provider packages. For instance, in Figure~\ref{fig:definitions_example}, \textsf{body-parser} is a provider of \textsf{express}; \textsf{body-parser} also has \textsf{bytes} as a provider. In this scenario, we consider \textsf{body-parser} to be a transitive provider.

    \item \textbf{client package release} is the package release that uses features and resources exposed by provider package releases. In Figure \ref{fig:definitions_example}, \textsf{express} is a client of \textsf{body-parser}, \textsf{body-parser} is a client of \textsf{bytes}, and so on.

    \item \textbf{direct provider release} is the one directly used by its client, that is, the package that the client explicitly declares as a dependency. In Figure \ref{fig:definitions_example}, \textsf{express} is a direct provider of \textsf{ember-cli}, and \textsf{bytes} is a direct provider of \textsf{body-parser}.

    \item \textbf{indirect provider release} is a package release that at least one of its providers uses. In other words, it is a provider of at least one of the direct client's providers. In Figure \ref{fig:definitions_example}, both \textsf{body-parser} and \textsf{bytes} are indirect providers of \textsf{ember-cli}, and \textsf{bytes} is an indirect provider of \textsf{express}.

    \item \textbf{transitive provider release} is the package release between the one that introduced a breaking change and the client. For example, if a breaking change is introduced by \textsf{bytes}, in Figure \ref{fig:definitions_example}, and affects client \textsf{ember-cli}, both packages \textsf{express} and \textsf{body-parser} are transitive providers. This is because the breaking change transited through these packages (\textsf{body-parser} and \textsf{express}) to arrive at client \textsf{ember-cli}. The transitive providers are all also impacted by the breaking change.

    \item \textbf{version statement}: a client can specify its provider's versions on \textit{package.json}, a metadata file used by \textsf{npm} to specify providers and their versions, among other purposes. The version statement contains the accepted version of a provider. For example, the \textit{version statement} in the following metadata \texttt{\{"dependencies": \{"express": "\textasciicircum4.10.6"\}\}}, defines that the client requires \textsf{express} on version \textsf{\textasciicircum4.10.6}.
    
    \item \textbf{version range}: on the \textit{version statement} a client can specify a range of versions/releases accepted by its provider. There are three types of ranges:

        \begin{itemize}
            \item \textbf{all ($>$=, or \textit{*})}: using this range, the client specifies that all new provider releases are supported/accepted and downloadable, even the ones with breaking changes.

            \item \textbf{caret (\textbf{\textasciicircum})}: with this range, the client specifies that all new provider releases that contain new features and bug fixes are supported/accepted and downloadable; breaking changes must be avoided. This is the default range used by \textsf{npm} when a dependency is installed.

            \item \textbf{tilde range (\texttildelow)}: this range specifies that all new provider releases that only contain bug fixes are supported/accepted and downloadable; breaking changes and new features must be avoided.

            \item \textbf{steady range}: this range always resolves to a specific version and is also known as \textit{specific range}. That is, the versioning statement has no range on it but rather a specific version. \textsf{npm} allows installation with a steady range using the command line option \textsf{--save-exact}.
        \end{itemize}

    \item \textbf{implicit and explicit update}: an \textit{implicit update} happens when the client receives a new provider version due to the range version in the \textit{package.json}. For a version statement defined with a range of versions, for example, \textsf{\textasciicircum4.10.6}, an implicit update happens when \textsf{npm} installs a version 4.10.9 that matches the range. An \textit{explicit update} takes place when the client manually updates the versioning statement directly in the \textit{package.json}.

\item \textbf{manifesting breaking changes} are provider changes that manifest as a fault on the client package, ultimately breaking the client's build. The adopted definition of breaking change by the prior literature~\cite{how_to_break_an_api,TosemBreaking,when_it_breaks,why_how_java,what-dependencies-tell-semver,bc_maven,using_others_tests,noregrets2018} includes cases that are not considered breaking changes (e.g., a change in an API that is not effectively used by a client package). Conversely, manifesting breaking changes include cases that are not covered by the prior definitions of breaking change (e.g., because the provider package is used in a way that is not intended by the provider developer, a semantic-version compliant change introduced by a new release of this provider causes an expected error in the client package).
\end{itemize}

\subsection{Motivating Examples}

We found the following two examples of manifesting breaking changes in our manual analysis (on each of the following Listing, red lines have been removed from the source code whereas blue lines have been inserted into the source code). Our manual analysis (Section~\ref{subsec:rq1}) consists of executing the client tests suite for its releases and analyzing all executions that run into an error.

The client \textsf{assetgraph-builder@7.0.0} has a provider \textsf{assetgraph@6.0.0} that has a provider \textsf{terser@\textasciicircum4.0.0}, but, due to a range of versions, \textsf{npm} installed \textsf{terser@4.6.10}. Release \textsf{4.3.0} of \textsf{terser} introduces a change which, by default, enables the wrapping of functions on parsing, as Listing \ref{cod:diff:terser}.\footnote{https://github.com/terser/terser/compare/v4.2.1..v4.3.0}

\begin{lstlisting}[numbers=none, language=diff, label=cod:diff:terser, caption={Diff between \textsf{terser@4.2.1} and \textsf{terser@4.3.0} default behavior.}]
   // terser@4.2.1 without default wrapping behavior
   foo(function(){});

   // terser@4.3.0 default wrapping behavior
   foo((function(){}));

\end{lstlisting}

This change breaks the \textsf{assetgraph-builder@7.0.0}'s tests.\footnote{https://github.com/terser/terser/issues/496} Once this feature is turned a default behavior, the client \textsf{assetgraph-builder@8.0.0} adopts its test to make it compatible with the \textsf{terser}'s behavior, as Listing \ref{cod:diff:assetgraph}.\footnote{https://github.com/assetgraph/assetgraph-builder/commit/e4140416e7feaa3d088cf3ad0229fd677ff36dbc}

\begin{lstlisting}[numbers=none, language=diff, label=cod:diff:assetgraph, caption={Diff with \textsf{assetgraph@8.0.0} client's tests adjusting to breaking change.}]
   expect(
     javaScriptAssets[0].text,
     'to match',
-     /SockJS=[\s\S]*define\("main",function\(\)\{\}\);/
+     /SockJS=[\s\S]*define\("main",\(?function\(\)\{\}\)   ?\);/
   );
\end{lstlisting}

\begin{figure}[H]
    \centering
    \includegraphics[scale=0.5]{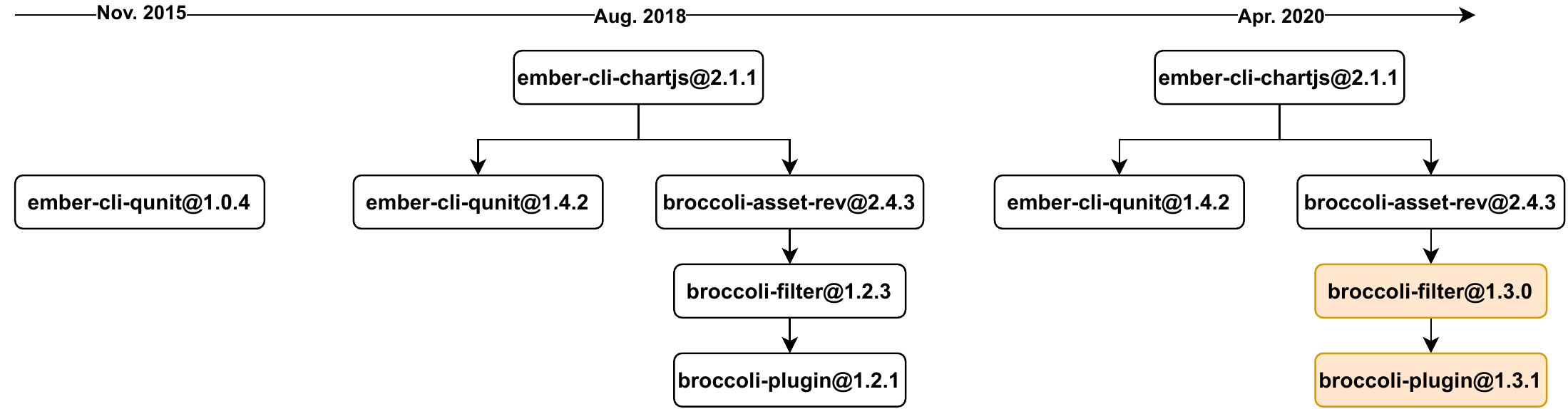}
    \caption{The evolution of the dependency tree (resolved versions) for \textsf{ember-cli-chartjs@2.1.1} when it was published (middle-tree) and when the associated tests with the release were executed in our study (right-hand tree).}
    \label{fig:bc_example_2}
\end{figure}

Sometimes, provider changes can break a client long after their introduction. This occurred in the client package \textsf{ember-cli-chartjs@2.1.1}. In Figure \ref{fig:bc_example_2}, the release \textsf{1.0.4} of \textsf{ember-cli-qunit} (left-tree) introduced a change that did not lead to a breaking change. However, almost three years later, \textsf{ember-cli-qunit} was used together with the release \textsf{1.3.1} of the provider \textsf{broccoli-plugin} (middle-tree), and a breaking change manifested.

In November 2015, the provider \textsf{ember-cli-qunit@1.0.4} fixed an error in its code, changing the returned object type of function \textsf{lintTree},\footnote{https://github.com/ember-cli/ember-cli-qunit/commit/6fdfe7d} as shown in Listings \ref{cod:bc:ember}. Despite being a type change, it did not break the client when it was released, and this fix was retained in further releases of \textsf{ember-cli-qunit}.

\begin{lstlisting}[numbers=none, language=diff, label=cod:bc:ember, caption={ember-cli-qunit@1.0.4 object type change.}]
 lintTree: function(type, tree) {
   // Skip if useLintTree === false.
   if (this.options['ember-cli-qunit'] && ... ) {
-      return tree;
+      // Fakes an empty broccoli tree
+      return { inputTree: tree, rebuild: function() { return []; } };
   }
\end{lstlisting}

Almost three years later, on Aug. 2018, the provider \textsf{broccoli-plugin@1.3.1} was released (middle-tree in Figure \ref{fig:bc_example_2}) to fix a bug,\footnote{https://github.com/broccolijs/broccoli-plugin/commit/3f9a42b} as in Listing \ref{cod:bc:ember_2}.

\begin{lstlisting}[numbers=none, language=diff, label=cod:bc:ember_2, caption={broccoli-plugin@1.3.1 validation function enhanced.}]
 function isPossibleNode(node) {
-  return typeof node === 'string' ||
-    (node !== null && typeof node === 'object')
+  var type = typeof node;
+  if (node === null) {
+    return false;
+  } else if (type === 'string') {
  ...
+  } else {
+    return false;
+  }
\end{lstlisting}

The release \textsf{1.3.1} of the \textsf{broccoli-plugin} package experienced a manifesting breaking change due to a fix in the provider \textsf{ember-cli-qunit@1.0.4},\footnote{https://github.com/broccolijs/broccoli-merge-trees/issues/65} which was released almost three years prior. This manifesting breaking change occurred because the ember-cli-chartjs' dependency tree evolved over time due to the range versions, as shown in Figure \ref{fig:bc_example_2}, causing the break. When the package \textsf{ember-cli-chartjs@2.1.1} was installed on April 2020 (the date of our analysis), the installation failed due to the integration of \textsf{broccoli-plugin@1.3.1} changes into \textsf{ember-cli-qunit}. Fifteen days later, \textsf{ember-cli-qunit@1.4.3} fixed the issue when the \textsf{ember-cli-qunit}'s object type was changed again.\footnote{https://github.com/ember-cli/ember-cli-qunit/commit/59ca6ad} During the fifteen-day period when the manifesting breaking change remained unresolved, \textsf{broccoli-plugin} received about \textit{384k} downloads from \textsf{npm}. This scenario shows that even popular and mature projects can be affected by breaking changes. Although we recognize that the download count does not necessarily reflect the popularity of a package, we use this metric as an illustrative example of how many client packages might have been impacted by a provider package.


\section{Study Design}

This section describes how we collected our data (Section~\ref{subsec:data}) and the motivation and approach for each RQ (Section~\ref{subsec:motivation_approach}).

\subsection{Data Collection}
\label{subsec:data}

\subsubsection{Obtaining metadata from \textit{npm} packages}
\label{subsubsec:obtaining_npm_metadata}
  
\begin{figure}[H]
	\centering
	\includegraphics[scale=0.65]{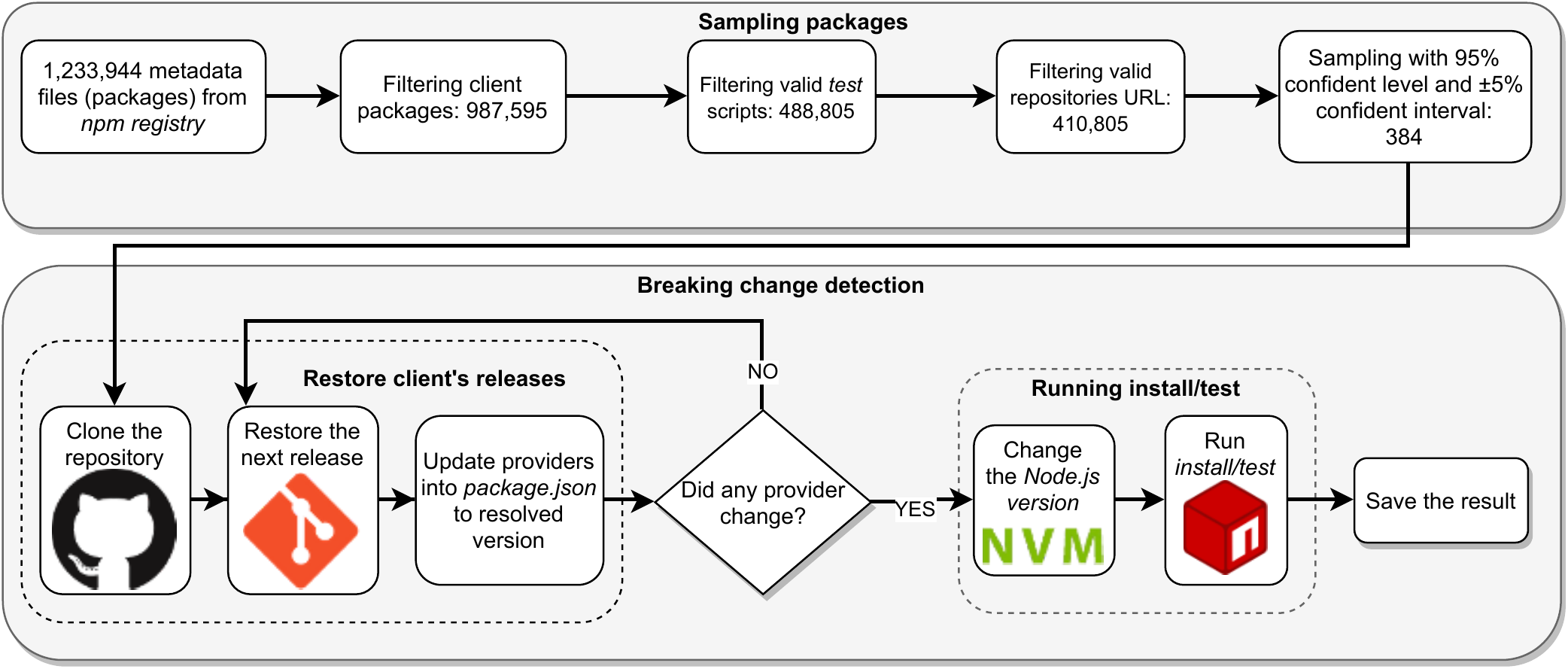}
	\caption{Approach to sampling the database and executing the associated tests with the client release.}
	\label{fig:bc_work}
\end{figure}

The first part of Figure~\ref{fig:bc_work} shows our approach for sampling the database. We initially gathered all the metadata files (i.e., \textit{package.json} files) from the published packages in the \textit{npm} registry between December 20, 2010 and April 01, 2020, accounting for 1,233,944 packages. This range refers to the oldest checkpoint that we could retrieve and the most recent one when we started this study. We ignored packages that did not have any providers in the \textit{package.json} since they cannot be considered client packages and will therefore not suffer breaking changes. After filtering packages without a provider, our dataset comprises 987,595 \textit{package.json} metadata files. For each release of each package, we recorded the timestamp of the release and the name of the providers with their respective versioning statements.

We parsed all the versioning statements and determined the resolved provider version at the time of each client release. Prior works have adopted similar approaches when studying dependency management~\cite{downgrades_by_filipe,technical_lag_tom_mens}. For each provider in each client release, we retrieved the most recent provider version that satisfied the range specified by the client in that release; i.e., the \textit{resolved version}. Using this resolved version, we determined whether a provider changed its version between the two client releases. In other words, we reproduced the adopted versions of all providers by \textit{resolving} the provider version at the release time of the client.

To further refine our sample, we analyzed two criteria in the associated \textit{package.json} snapshot with the latest version of the client packages in our dataset: 

\begin{enumerate}
    \item The \textit{package.json} snapshot should have a non-empty entry for the ``script test'' field, and the entry should differ from the default: \texttt{Error: no test specified}. We specified this criterion in order to run the automated tests that were part of our method to detect manifesting breaking changes. In total, 488,805 packages remained after applying this criterion.
    
    \item The \textit{package.json} snapshot should have an entry containing the package's repository \texttt{URL}, as we wanted to retrieve information from the package codebase. After applying this criterion, 410,433 packages remained in our dataset.
\end{enumerate}

{}

\subsubsection{Running clients' tests}

\label{subsubsec:identifying_breaking_changes}
Given the size of our dataset (more than 410,000 client packages), we ran tests on a random sample. At a 95\% confidence level and $\pm$5\% confidence interval, we randomly selected 384 packages. Our sample has a median of 5.5 releases and 9 direct providers per package. We chose to study a random sample since our manual analysis is slow to run over a large dataset (Section \ref{subsection:sanity-check-failure-cases}); we spent a month executing our method in our sample. We did not ignore packages based on the number of releases or providers or any other metric. We performed a manual check on all selected packages that had fewer than four releases (130 out of 384) by checking their repositories and aiming to remove packages that are not real projects, lack tests, lack code, are example projects, etc. When we removed one package, we sampled another one following the two criteria described above. 

The second part of Figure~\ref{fig:bc_work} depicts our approach to running the test scripts for each release of the 384 clients. For each client package, we cloned its repository -- all client repositories are hosted on \textsf{GitHub} -- and restored the work tree of all releases using their respective release tags (e.g., ``v1.0.0''). For releases that are not tagged, we used their provided timestamp in the \textit{package.json} metadata to restore the work tree (i.e., we matched the release timestamp and the closest existing commit in the master branch). We conducted an analysis and verified that tags and timestamp point to the same commit in 94\% of releases with tags, thus checkout based on timestamps is reliable for untagged releases.

After restoring the work tree of a client release, we updated all versioning statements in the associated \textit{package.json} entry with the specific resolved provider version (see Section~\ref{subsubsec:obtaining_npm_metadata}). We then excluded a file called \textit{package-lock.json}, which \textit{locks} the providers and indirect providers versions. We also executed the associated tests on a release of the client package whenever a provider package changed in that release, as this can potentially introduce a manifesting breaking change. A provider change can be: 1) a provider added into the \textit{package.json}; or 2) the resolved version of a provider changed between the previous and current release of the client package. 

We sought to reproduce the same build environment that existed when the provider changed. Therefore, before executing the tests of the client packages, we performed a best-effort procedure to identify the \textsf{Node.js} that was adopted by the client package at the time the provider changed. This was because every six months a new major version of \textsf{Node.js} is released. \footnote{https://github.com/nodejs/node\#release-types} As we wanted to reproduce the test results with respect to the time when the client package published its release, we changed the \textsf{Node.js} version before executing the client package tests. We selected the \textsf{Node.js} version using two different approaches. Our preferred approach was to select the same \textsf{Node.js} version as the one specified in the \texttt{engines$\rightarrow$node} field of the \textit{package.json} file.\footnote{https://docs.npmjs.com/files/package.json\#engines} This field allows developers to manually specify the \textsf{Node.js} version that runs the associated code with the build of a specific release. When this field was not set, we selected the latest \textsf{Node.js} version available\footnote{https://nodejs.org/en/download/releases} at the time of the client package release. Therefore, we changed the \textsf{Node.js} version, executed the install script, and released tests using the \texttt{npm install} and \texttt{npm test} commands, respectively. If the install or test commands failed due to incompatibilities with the selected \textsf{Node.js} version -- or took more than 10 minutes --, we changed to the previous major release of \textsf{Node.js} until the install and test commands succeeded. We used the Node Version Manager (\textsf{NVM}) tool to exchange \textsf{Node.js} versions.
Additionally, we also changed the \textsf{npm} version according to the \textsf{Node.js} version. \textsf{npm} is the package manager to \textsf{Node.js} packages and executes the \texttt{install} and \texttt{test} scripts. We performed the same procedure to select the \textsf{npm} version to use during the installation and test runs. Finally, we executed the install/test scripts and saved the results (success or error) for each client release.

After executing the install/test scripts of the 384 client packages in our sample, we discarded 33 packages because the errors did not allow the execution of the install/test script in any of their releases: 15 clients did not have one of the required files; 11 had invalid test scripts (e.g., \texttt{\{"test": "no test"\}}); 4 listed some required files in the \textit{.gitignore} file -- that specifies untracked files that \textsf{git} should ignore;\footnote{https://git-scm.com/docs/gitignore} 2 required specific database configurations that could not be done; and 1 package required a \textit{key} to access a server. We randomly replaced these 33 packages following the aforementioned criteria.

Table \ref{tab:data_execution} shows the results of the execution of the install/test scripts of the 384 client packages and their 3,230 releases. Since the associated providers' version with 2,727 releases did not change, these tests' releases were not executed. Finally, we consider as possible manifesting breaking changes cases in which all client packages and releases failed the install/test scripts.

\begin{table}[H]
	\centering
	\caption{Results of execution of the install/test scripts.}
	\begin{tabular}{lrr}
		\toprule
		          Tests      & Client Packages & Releases \\ \hline
		
		Executed        & 384     & 3,230     \\
		Not executed    & 0       & 2,727     \\
		Success         & 181     & 1,954     \\
		Fails           & 203     & 1,276     \\
		\midrule
		Total           & 384     & 5,957     \\
		\bottomrule
	\end{tabular}
	\label{tab:data_execution}
\end{table}

A replication package including our client packages sample, instruments, scripts, and identified manifesting breaking changes is available for download at \url{https://doi.org/10.5281/zenodo.5558085}.

\subsubsection{Manual check on failure cases: detecting manifesting breaking changes} 
\label{subsection:sanity-check-failure-cases}
For all failure cases (203 clients and 1,276 releases) on the execution of install/test scripts, we manually analyzed which ones were true cases of manifesting breaking changes. To identify breaking changes that manifest themselves in a client package, we leveraged the output logs (logs generated by \textsf{npm} when executing the \texttt{install} and \texttt{test} scripts) generated as the result of executing the method described in Section~\ref{subsubsec:identifying_breaking_changes} (see the second part of Figure~\ref{fig:bc_work}). For each failed test result, we obtained the error description and the associated stack trace. We then differentiated failed test results caused by a related issue with the client package (e.g., an introduced bug by the client) from those caused by a change in the provider package (e.g., a change in the return type of a provider's function). From the obtained stack traces, we determined whether any function of a provider package was called and manually investigated the positive cases. During our manual investigation, we sought to confirm that the test failure was caused by a manifesting breaking change introduced by the provider package. 

The first author was responsible for running the tests and identifying the manifesting breaking changes and related releases and commits. The first author also manually analyzed each of the manifesting breaking changes and recorded the following information about each of them: the number of affected versions of the client; whether any documentation mentions the manifesting breaking change; the responsible package for addressing the breaking change (provider or client); the client version impacted by the manifesting breaking change; the provider version that introduced the breaking change; and a textual description about the causes for the breaking change manifestation (e.g., "the provider function was renamed by mistake", "The provider normalizeurl@1.0.0 introduce[d] a new function and the client assetgraph use[d] it. But the client forgot to update the provider version in package.json.", "The provider inserts a " " in a null body request"). During this process, several rounds of discussions were performed among the authors to refine the analysis, using continuous comparison \cite{strauss1998basics} and negotiated agreement \cite{garrison2006revisiting}. In the negotiated agreement process, the researchers discussed the rationale they used to categorize each code until reaching consensus \cite{garrison2006revisiting}. More specifically, we leveraged the recorded information about each manifesting breaking change to derive a consistent categorization of the introduced breaking changes (RQ2 and RQ3) and to guide new iterations of the manual analysis.

More specifically, the following set of actions was performed during our manual investigation:

\begin{itemize}
	\item \textbf{Analyze the execution flow:} To determine whether the associated function with the test failure occurred in the provider or the client code, we leveraged the stack traces to identify which function was called when the test failed. In particular, we instrumented the code of the provider and the client packages to output any necessary information to analyze the execution flow. We analyzed the variable contents by adding a call to the \textsf{console.log()} and \textsf{console.trace()} functions in each part of the code where the client package calls a function of the provider. For example, suppose the following error appeared: \textit{``TypeError: myObject.callback is not a function''}.To discover the variable content, we use the command \texttt{console.log(myObject)} to check whether myObject variable was changed, null, or received other values.
	
	\item \textbf{Analyze the status of the Continuous Integration (CI) pipeline:} We compared the status of the CI pipeline between the originally built release and the status of CI pipeline at the time of our manual investigation. Since the source code of the client package remains the same between the original release and the installed version in our analysis, we use the difference between the status of the CI pipeline as additional evidence that the test failure was caused by a provider version change. Not all clients had CI pipelines, but when they had, it was helpful.
	
	\item \textbf{Search for client fixing commits:} We manually searched for recovering commits in the history of commits between the installed and previous releases of the client package. Whenever a recovery commit was identified (by reading the commit message), we determined whether the error was due to the client or the provider code. For example, we observed cases in which a client updated a provider in the release with failed tests. We also observed that, in the following commits, the provider was downgraded and the commit message was \textit{``downgrade provider''} or \textit{``fix breaking change''}. In these cases, we considered the test failure as caused by a manifesting breaking change.
	
	\item \textbf{Search for related issue reports and pull requests:} We hypothesized that a manifesting breaking change would affect different clients that, in turn, would either issue a bug report or perform a fix followed by a pull request to the codebase of the provider package. Therefore, we searched for issue reports and pull requests with the same error message obtained in our stack trace. We then collected detailed information about the error to confirm whether it was due to a manifesting breaking change introduced by the provider package.
	
	\item \textbf{Previous and subsequent provider versions:} If the test error was caused by a manifesting breaking change, downgrading to the previous provider version or upgrading to a subsequent provider version might fix the error, if the provider already fixed it. \textit{Subsequent provider versions} means all provider versions that fit the versioning statement and are greater than the provider version that introduced the manifesting breaking change (i.e., the adopted provider version when the test failed). In this case, we uninstalled the current version and installed the previous and subsequent versions, and executed the test scripts again. For example, if the client specified a provider \textsf{p} as \texttt{\{"p": "\textasciicircum1.0.2"\}} that brought about a breaking change in the version, for example, \textsf{1.0.4}, we installed \textsf{p@1.0.2}, \textsf{p@1.0.3}, and \textsf{p@1.0.5} to verify whether the error persisted for those versions.
\end{itemize}

\subsection{Research questions: motivation, approach}
\label{subsec:motivation_approach}

This section contains the motivation and the approach for each of the research questions.

\noindent \subsubsection{\textbf{RQ1. To what extent do manifesting breaking changes manifest in client packages?}}
\label{subsec:rq1}
\hfill\\

\smallskip \noindent 
\textbf{Motivation:} By default, \textsf{npm} sets the caret range as a default versioning statement that automatically updates minor and patch releases. Hence, manifesting breaking changes that are introduced in minor and patch releases can inadvertently cause downtime in packages that are downloaded hundreds of thousands of times per day, affecting a large body of software developers. Understanding the prevalence of manifesting breaking changes in popular software ecosystems such as \textsf{npm} is important to help developers assess the risks of accepting automatic minor and patch updates. Although prior studies have focused on the frequency of API breaking changes~\cite{how_to_break_an_api}, breaking changes can occur for different reasons. Determining the prevalence of a broader range of breaking change types remains an open research problem.

\noindent
\textbf{Approach:} 
For all cases that resulted in an error on the install/test script, we determined the type of error (client, provider, not discovered). We calculated, out of the 384 packages and 3,230 releases, the percentage of cases that we confirmed as manifesting breaking change. Considering all the providers on the client's latest releases, we calculated the percentage of providers that introduced manifesting breaking changes. In addition, we calculated how many times (number of releases) each provider introduced at least one manifesting breaking change.

\noindent 
\subsubsection{\textbf{RQ2. What problems in the provider package cause a manifesting breaking change?}}
\label{sub:rq2}
\hfill\\

\textbf{Motivation:} 
Prior studies about breaking changes in the \textsf{npm} ecosystem are restricted to APIs' breaking changes \cite{detecting_bc_JavaScript_apis}. However, other issues that provider packages can introduce in minor and patch releases can manifest a breaking change. To support developers to reason about manifesting breaking changes, it is important to understand their root causes.

\noindent
\textbf{Approach:} 
In this RQ, we analyzed the type of changes introduced by provider packages that bring about a manifesting breaking change. With the name and version of the provider packages, we manually analyzed the provider's repository to find the exact change that caused a break. We used the following approaches to find the specific changes introduced by providers:

\begin{itemize}
	\item \textbf{Using diff tools:} We used diff tools to analyze the introduced change between two releases of a provider. For example, suppose that a manifesting breaking change was introduced in the release \textsf{provider@1.2.5}. In this case, we retrieved the source code of previous versions, e.g., \textsf{provider@1.2.4}, and performed the diff between these versions to manually inspect the changed code.
	
	\item \textbf{Analyzing provider's commits:} We used the provider's commits to analyze the changes between releases. For a manifesting breaking change in the provider \textsf{p}, we verified its repository and manually analyzed the commits ahead or behind the release tag commit that introduced a manifesting breaking change.
	
	\item \textbf{Analyzing changelogs:} Changelogs contain information on all relevant changes in the history of a package. We used these changelogs to understand the introduced changes in a release of a client package and to verify whether any manifesting breaking change fix was described.
\end{itemize}

We also looked at issue reports and pull requests for explanations of the causes of manifesting breaking changes. After discovering the provider changes that introduced breaking changes, we analyzed, categorized, and grouped common issues. For example, all related issues to changing object types were grouped into a category called \textit{Object type changed}. Furthermore, we analyzed the Semantic Version level that introduced and fixed/recovered the manifesting breaking changes both in the provider and client packages to verify the relationship between manifesting breaking changes and non-major releases.

We analyzed the version numbering of releases that fixed a manifesting breaking change and where manifesting breaking changes were documented (changelogs, issue reports, etc.). Furthermore, we analyzed the depth of the dependency tree of the provider that introduced a manifesting breaking change, since 25\% of npm packages had at least 95 transitive dependencies in 2016~\cite{npm-three}.

\noindent \subsubsection{\textbf{RQ3. How do client packages recover from a manifesting breaking change?}}
\hfill\\
\label{sub:rq3}
\noindent

\textbf{Motivation:} 
A breaking change may impact the client package through an \textit{implicit} or \textit{explicit} update. A client recovery is identified by an update to its code, by waiting for a new provider's release, or by performing a downgrade/upgrade in the provider's version. Breaking changes may be caused either by a \textit{direct} or \textit{indirect} provider since the client packages depend on a few direct providers and many indirect ones~\cite{npm-seven}. A breaking change may cascade to transitive dependencies if it remains unfixed. Even if the client packages can recover from the breaking change by upgrading to a newer version of the provider package, the client packages can manually resolve incompatibilities that might exist \cite{efficient_static_checking}. Understanding how breaking changes manifest in client packages can help developers understand how to recover from them.
\noindent

\textbf{Approach:} 
We retrieved all information for this RQ from the clients' repositories. We searched for information about the error and how the client packages recovered from the manifesting breaking change. The following information was analyzed:

\begin{itemize}
	\item \textbf{Commits:} We manually checked the subsequent commits of the client packages that were pushed to their repositories after the provider release that introduced the respective manifesting breaking change. In particular, we searched for commits that touched the \textit{package.json} file. In the file history, we checked if the provider was downgraded, upgraded, replaced, or removed.
	
	\item \textbf{Changelogs:} We analyzed the client changelogs and release notes looking for mentions of provider updates/downgrades. About 48\% of clients maintained a changelog or release notes in their repositories.
	
	\item \textbf{Pull requests/Issue reports:} We searched for pull requests and issue reports in the client repository that contained information about the manifesting breaking changes. For example, we found pull requests and issue reports with ``Update provider'' and ``Fix provider error'' in the title.
\end{itemize}

For each manifesting breaking change case, we recovered the provider's dependency tree. For example, in our second motivating example (Section~\ref{sec:motivating}), we recovered the dependency tree from the client to the package that introduced the manifesting breaking change, which resulted in \textsf{broccoli-asset-rev$\rightarrow$broccoli-filter$\rightarrow$broccoli-plugin} (Figure \ref{fig:bc_example_2}). We investigated how many breaking change cases were introduced by direct and indirect providers, when the manifesting breaking change was introduced and fixed/recovered, which package fixed/recovered from it, and how it was fixed/recovered. We also verified how client packages changed the provider's versions and how the associated documentation with manifesting breaking changes related to the time to fix it.

\subsection{Scope and Limitations}

As our definition of manifesting breaking changes includes cases that are not included by the prior definitions of breaking changes (see Section~\ref{subsec:definitions}), this paper does not intend to provide a direct comparison between these two phenomena. As a result, the stated research questions do not indicate the proportion of manifest breaking changes that are, in fact, breaking changes as defined by prior literature (e.g., an API change by the provider). In addition, since provider packages are rarely accompanied by any formal specification of their intended behavior, it is impossible at the scale of our study to differentiate errors that manifest in the client package due to breaking changes from those that manifest due to an idiosyncratic usage of the provider by the client package. Therefore, the results of the stated RQs cannot be used to assess whether a client package could fix its build by simply updating to a newer version of the provider.


\section{Results}
\label{sec:results}

This section presents the associated findings for each RQ.


\subsection{RQ1. How often do manifesting breaking changes occur in the client package?}\label{subsec:rq1F}
\noindent

\textbf{Finding 1:}
\textit{\textbf{11.7\% of the client packages (regardless of their releases) and 13.9\% of the client releases were impacted by a manifesting breaking change.}} 
From all 384 client packages, 45 (11.7\%) suffered a failing test from a manifesting breaking change in at least one release. From 3,230 client releases for which the tests were executed, 1,276 failed, and all errors were manually analyzed. In 450 (13.9\%) releases, the error was raised by the provider packages, characterizing a manifesting breaking change. On 86 (2.7\%) releases, we could not identify which package raised the error.

We detected that 261 (8.1\%) releases suffered a particular error type that we call \textit{breaking due to external change}. These releases used a provider that relied on data/resources from an external API/service (e.g., \textsf{Twitter}) that were no longer available, impacting all client's releases. The provider cannot fix this error, because it does not own the resource. These cases imply that detecting manifest breaking changes by running the clients’ tests can introduce false positives, which we simply ignored during our manual analyses. We also considered cases in which a provider package was removed from \textsf{npm} as \textit{breaking due to external change}. Table \ref{tab:releases_analyses} shows the results of analyses by releases.

\begin{table}[H]
	\centering
	\caption{Results of releases' analyses.}
	\begin{tabular}{llcc}
		\toprule
		\textbf{Results}        & \phantom{ab} &\textbf{Releases (\#)} & \textbf{(\%)} \\ \hline
		Success                 & \phantom{ab} & 1954         & 60.5          \\
		\midrule
		\multirow{4}{*}{Fail} 
		& Client's errors         & 479 & 14.8 \\
		& manifesting breaking changes        & 450 & 13.9 \\
		& Breaking due to external changes & 261 & 8.1  \\
		& Errors not identified   & 86  & 2.7  \\
		\midrule
		\multicolumn{2}{l}{Total} & \textbf{3230} & \textbf{100} \\
		\bottomrule
	\end{tabular}
	\label{tab:releases_analyses}
\end{table}

\noindent

\textbf{Finding 2: \textit{92.2\% providers introduced a single manifesting breaking change.}} In our sample, 47 providers (92.2\%) of 51 introduced a single release with a manifesting breaking change, and four providers introduced two releases with manifesting breaking changes. We detected 55 unique manifesting breaking change cases introduced by providers, some of which impacted multiple clients. For example, the breaking change exhibited in the \textit{Incompatible Providers Versions} classification (Finding 3) impacted six clients. Therefore, 64 manifesting breaking change cases manifested in the client packages. Finally, there were 1,909 providers on all clients' latest versions, and the percentage of providers that introduced manifesting breaking change was 2.6\% (51 of 1909).

\vspace{5pt}
\begin{mdframed}
\begin{itemize}
    \item About 11.7\% of clients and 13.9\% of their releases suffered from manifesting breaking changes.
    \item We detected failing tests due to 2\% of the providers with changes.
    \item Over 90\% of those that introduced manifesting breaking changes did so through just a single release with a manifesting breaking change.
\end{itemize}
\end{mdframed}


\subsection{RQ2. What issues in the provider package caused a breaking change to manifest?}
\noindent
\textbf{Finding 3: \textit{We found 8 categories of {issues}}}. 
We grouped each manifesting breaking change into eight categories, depending on its root cause (issue). Table~\ref{tab:bc_category} presents each category, the number of occurrences, and the number of impacted client releases.

\begin{table}[H]
	\centering
	\caption{The identified categories of manifesting breaking changes.}
	\begin{tabular}{lrrrrr} \toprule
		\textbf{Category} & \multicolumn{2}{c}{\textbf{Cases}} & \phantom{ab} & \multicolumn{2}{c}{\textbf{Releases}}
		\\
		\cmidrule{2-3} \cmidrule{5-6}
		& (\#) & (\%) && (\#) & (\%) \\ \midrule
		Feature change                  & 25 & 39.1  && 101   & 22.4                \\
		Incompatible providers versions & 15 & 23.4  && 64    & 14.2                \\
		Object type changed             & 9  & 14.1  && 213   & 47.3                \\
		Undefined object                & 5  & 7.8   && 28    & 6.2                 \\
		Semantically wrong code                      & 5  & 7.8   && 14    & 3.1                 \\
		Failed provider update          & 2  & 3.1   && 24    & 5.3                 \\
		Renamed function                & 2  & 3.1   && 2     & 0.4                 \\
		File not found                  & 1  & 1.6   && 4     & 0.9                 \\ \hline
		\textbf{Total}                  & \textbf{64} &&       & \textbf{450} &       \\
		\bottomrule
	\end{tabular}
	\label{tab:bc_category}
\end{table}

In the following, we describe each category and present an example that we found during our manual analysis.

\begin{itemize}
   \item \textbf{Feature change:} manifesting breaking changes in this category are related to modifications of provider features (e.g., the default value of variables). An example happens in \textsf{request@2.17.0} -- this version was removed from \textsf{npm}, but the introduced change remained in the package -- when developers introduced a new decision rule into their code\footnote{https://github.com/request/request/commit/d05b6ba} as shown in Listing \ref{diff:bc_category_change_rule_1}.

\begin{lstlisting}[numbers=none, language=diff, label=diff:bc_category_change_rule_1, caption={Example of a manifesting breaking change categorized as feature change.}]
  debug('emitting complete', self.uri.href)
+ if(response.body == undefined && !self._json) {
+   response.body = "";
+ }
  self.emit('complete', response, response.body)
\end{lstlisting}

In Listing \ref{diff:bc_category_change_rule_1}, the provider \textsf{request} assigns an empty string to the \texttt{response.body} variable, instead of preserving \texttt{response.body} with its default \texttt{undefined} value.

   \item \textbf{Incompatible providers versions:} In this category, the client breaks because of a change in an indirect provider. An example happens in the packages \textsf{babel-eslint} and \textsf{escope}, where \textsf{escope} is an \textit{indirect} provider of \textsf{babel-eslint}. 

    \begin{lstlisting}[numbers=none, language=diff, label=cod:bc_category_incompatibles_providers, caption={Incompatible provider's versions example.}]
  }
-   },
-   visitClass: {
+ }, {
+   key: 'visitClass',
        value: function visitClass(node) {
    \end{lstlisting}

	The release \textsf{escope@3.4} introduced the presented change in Listing \ref{cod:bc_category_incompatibles_providers}. This change impacted the package \textsf{babel-eslint},\footnote{https://github.com/babel/babel-eslint/issues/243} even though the \textsf{escope} had not been a direct provider to \textsf{babel-eslint}.\footnote{https://github.com/estools/escope/issues/99\#issuecomment-178151491} This manifesting breaking change remained unresolved for a single day, during which \textsf{babel-eslint} received about \textit{80k} downloads from \textsf{npm}.
	
   \item \textbf{Object type changed:} 
	We detected 9 (14.06\%) cases in which the provider changed the type of an object, resulting in a breaking change in the client packages.

    \begin{lstlisting}[numbers=none, language=diff, label=cod:bc_category_change_type, caption={Object type changed example.}]
  this.setup();
- this.sockets = [];
+ this.sockets = {};
  this.nsps = {};
    this.connect Buffer = [];
  }
  var socket = nsp.add(this, function() {
-   self.sockets.push(socket);
+   self.sockets[socket.id] = socket;
    self.nsps[nsp.name] = socket;
    \end{lstlisting}

	In Listing \ref{cod:bc_category_change_type}, the provider \textsf{socket.io@1.4.0} turned an array into an object,\footnote{https://github.com/socketio/socket.io/commit/b73d9be}. This simple change broke many of \textsf{socket.io}'s clients, even the package \textsf{karma},\footnote{https://github.com/socketio/socket.io/issues/2368} a browser test runner, which was forced to update its code\footnote{https://github.com/karma-runner/karma/commit/3ab78d6} and publish \textsf{karma@0.13.19}. During the single day, the manifesting breaking change remained unresolved, \textsf{karma} was downloaded about \textit{146k} times from \textsf{npm}.

   \item \textbf{Undefined object:} In this category, an undefined object causes a runtime exception that breaks the provider, which throws the exception to the client package.

    \begin{lstlisting}[numbers=none, language=diff, label=cod:bc_category_undefined_object, caption={Undefined object code example.}]
+ app.options = app.options || {};
  app.options.babel = app.options.babel || {};
  app.options.babel.plugins = app.options.babel.plugins || [];
    \end{lstlisting}

	This error happened in the provider \textsf{ember-cli-htmlbars-inline-precompile@0.1.3}, which solved it as shown in Listing \ref{cod:bc_category_undefined_object}\footnote{https://github.com/ember-cli/ember-cli-htmlbars-inline-precompile/pull/5/commits/b3faf95}.

   \item \textbf{Failed provider update:} 
	In this category, provider \textit{A} updates its provider \textit{B}, but provider \textit{A} does not update its code to work with the new provider \textit{B}. We detected two cases of this category. In addition to an explicit update, one provider \textit{A} from this category specified its provider \textit{B} as an \textit{accept-all} range ($>$=). Over time, its provider \textit{B} published a major release that introduced a manifesting breaking change. Despite provider \textit{A} specifying an \textit{accept all} range, it did not consider the implicit update of provider \textit{B} and the client suffered an error.

   \item \textbf{Semantically wrong code:} manifesting
  breaking changes in this category happen when the provider writes a semantically wrong code, generating an error in its \textit{runtime process}\footnote{https://hacks.mozilla.org/2017/02/a-crash-course-in-just-in-time-jit-compilers/} and affecting the client. These errors could be caught in compile-time in a compiled language, but in \textsf{JavaScript} these errors happen at runtime. This occurred in the provider \textsf{front-matter@0.2.0} and four other cases.

	 \begin{lstlisting}[numbers=none, language=diff, label=cod:bc_category_wrong_code, caption={Semantically wrong code example.}]
  const separators = [ '---', '= yaml =']
- const pattern = pattern = '^('
+ const pattern = '^('
    + '((= yaml =)|(---))'
	 \end{lstlisting}

	On Listing \ref{cod:bc_category_wrong_code}, the provider repeated the variable name (\textit{pattern}) on its declaration, which generated a semantic error. Although this error can be easily detected and fixed, as the provider did\footnote{https://github.com/jxson/front-matter/commit/f16fc01} in Listing \ref{cod:bc_category_wrong_code}, the provider took almost one year to fix it (\textsf{front-matter@0.2.2}). Meanwhile, \textsf{front-matter} received about \textit{366} downloads in that period.

   \item \textbf{Renamed function:} 
   The manifesting breaking changes in this category occur when functions are renamed. Our analysis revealed 2 cases in which the functions were renamed. The renaming case is our first motivating example (Section~\ref{sec:motivating}); we describe the second one below.
   
	 \begin{lstlisting}[numbers=none, language=diff, label=cod:bc_category_renamed_function, caption={Renamed function code example.}]
- RedisClient.prototype.send_command = function (command, args, callback) {
-     var args_copy, arg, prefix_keys;
+ RedisClient.prototype.internal_send_command = function (command, args, callback) {
+     var arg, prefix_keys;
	 \end{lstlisting}

	The provider \textsf{redis@2.6.0-1} renamed a function, as in Listing \ref{cod:bc_category_renamed_function}.\footnote{https://github.com/NodeRedis/node-redis/commit/861749f} However, this function was used in a client package \textsf{fakeredis},\footnote{https://github.com/NodeRedis/node-redis/issues/1030\#issuecomment-205379483}, which broke with this change. Client package \textsf{fakeredis@1.0.3} recovered from this error by downgrading to \textsf{redis@2.6.0-0}.\footnote{https://github.com/hdachev/fakeredis/commit/01d1e99} In the five days period within which the manifesting breaking change was not fixed, \textsf{fakeredis} received about \textit{2.3k} downloads from \textsf{npm}.

   \item \textbf{File not found:} 
   In the cases in this category, the provider removes a file or adds it to the version control ignore list (\textit{.gitignore}) and the client tries to access it. In the unique case of this category in our sample, the provider referenced a file that was added to the ignore list.
\end{itemize}

\noindent
\textbf{Finding 4: \textit{ manifesting breaking changes are often introduced in patch releases.}} As shown in Table \ref{tab:semver_levels}, of the 64 cases of manifesting breaking changes we analyzed, three cases were introduced in major releases, 26 in minor releases, 28 in patch releases, and 5 in pre-releases. Although we only analyzed manifesting breaking changes from minor and patch releases, in three cases the manifesting breaking changes were introduced at major levels in an indirect provider, which transitively affected client packages---as in the \textsf{jsdom@16} case (see Section~\ref{sec:motivating}). 

\begin{table}[H]
	\centering
	\caption{manifesting breaking changes in each Semantic Version level.}
	\begin{tabular}{lrr}
		\toprule
		\textbf{Levels} & \textbf{(\#)} & \textbf{(\%)} \\ \hline
		Major           & 3    & 4.7      \\
		Minor           & 28   & 43.75     \\
		Patch           & 28   & 43.75     \\
		Pre-release     & 5    & 7.8       \\ 
		\midrule
		Total & 64 & 100 \\
		\bottomrule
	\end{tabular}
	\label{tab:semver_levels}
\end{table}

Pre-releases precede a stable release and are considered unstable; anything may change until a stable version is released.\footnote{https://semver.org/\#spec-item-9} In all detected breaking changes in pre-releases, the providers introduced unstable changes in pre-releases and propagated these changes to stable versions. An example is the pre-release \textsf{redis@2.6.0-1} (described in Section~\ref{sub:rq2}), whose rename of a function propagated to the stable version and caused a failure in the client packages. 

\begin{figure}[H]
	\centering
	\includegraphics[scale=0.5]{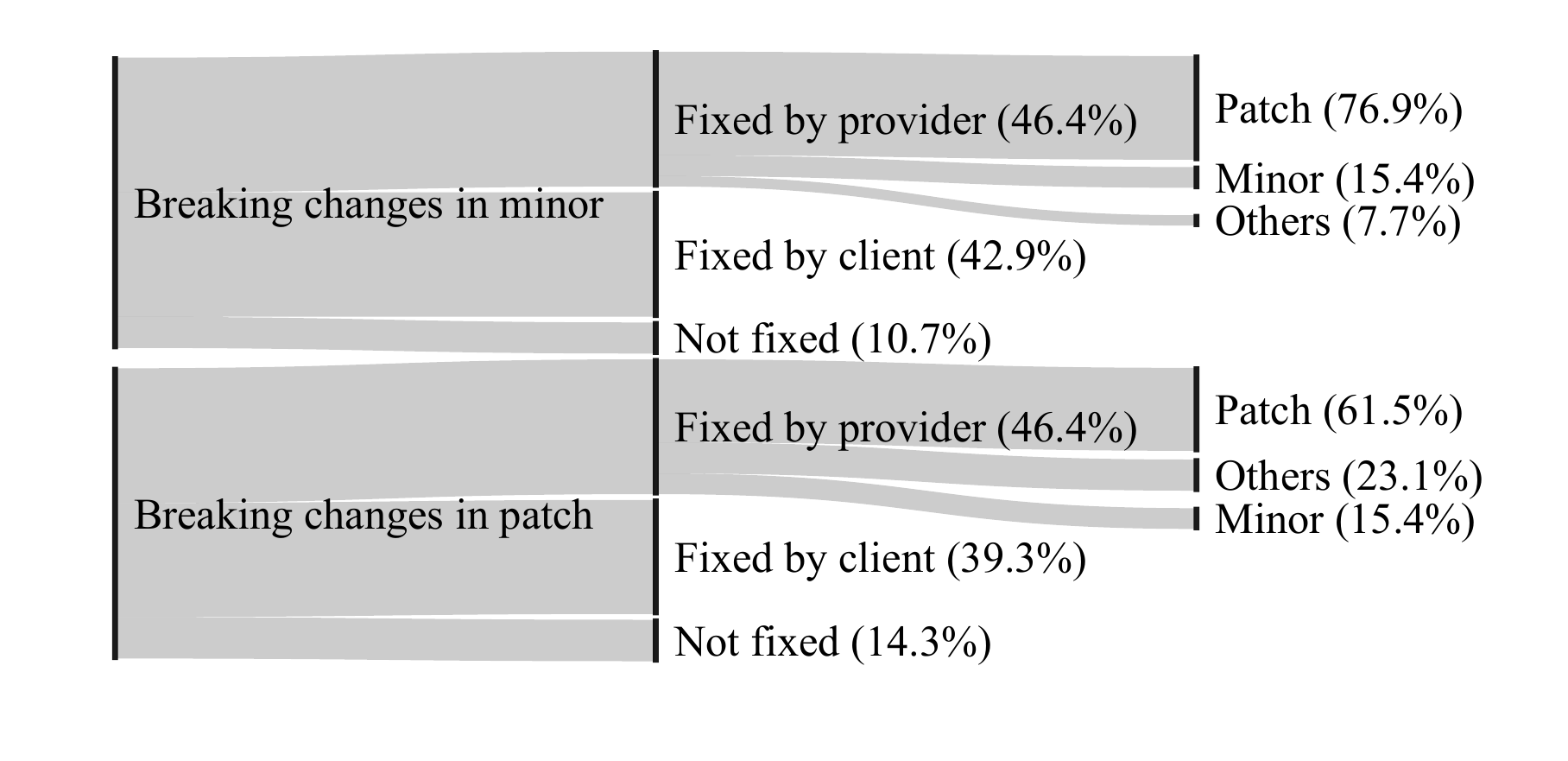}
	\caption{Proportion of fixed/recovered manifesting breaking changes by provider and client packages and the respective Semantic Version level of the fixing/recovering releases.}
	\label{fig:semver_fixed}
\end{figure}

\noindent
\textbf{Finding 5: \textit{manifesting breaking change fixes/recoveries are introduced by both clients and/or providers.}} 
We searched to identify which package fixed/recovered from the manifesting breaking changes -- client or provider -- and at which level the fixed/recovered release was published, as depicted in Figure~\ref{fig:semver_fixed}.

{}

Figure~\ref{fig:semver_fixed} shows that client packages recover from nearly half of the manifesting breaking change introduced in minor updates. In turn, 76.9\% of the manifesting breaking changes that are introduced by providers in a minor release are fixed in a patch release. Providers fix the majority of the manifesting breaking changes introduced in patch releases (46.4\% of the time), typically through a patch release (61.5\%).

\noindent
\textbf{Finding 6: \textit{21.9\% of the manifesting breaking changes are not documented.}}
Although clients and providers often document the occurrence or repair of a manifesting breaking change in issue reports, pull requests, or changelogs, more than one-fifth of the manifesting breaking changes are undocumented.

\begin{table}[H]
	\centering
	\caption{Summary of the proportion of documented manifesting breaking changes when they are introduced and fixed/recovered.}
	\begin{tabular}{lrrr}
		\toprule
		\textbf{Documentation} & \textbf{Introduced} & \textbf{Fixed/recovered} & \textbf{Proportion (\%)} \\ \hline
		Issue report                 & --             & 32           & 64            \\
		Pull request           & 5              & 15           & 44            \\
		Changelog              & 23             & 16           & 78   \\ 
		\bottomrule
	\end{tabular}
	\label{tab:bc_documentation}
\end{table}

Table \ref{tab:bc_documentation} shows that client and provider packages documented manifesting breaking changes in 78.1\% of all manifesting breaking changes. Out of all cases that have documentation, 70\% have more than one type of documentation. For example, the provider received an issue report, fixed the manifesting breaking change, and documented it in a changelog. Documenting manifesting breaking changes and their fixes supports client recovery (Section~\ref{sub:rq3}). 

\noindent
\textbf{Finding 7: \textit{57.8\% of the manifesting breaking changes are introduced by an indirect provider.}}
Indirect providers might also introduce manifesting breaking changes, which can then propagate to the client. Table \ref{tab:dependency_tree_deep} shows the depth level in the dependency tree of each provider that introduced a manifesting breaking change. About 42.2\% of manifesting breaking changes are introduced by a direct provider in the client's \textit{package.json}. These providers are the ones the client directly installs and that perform function calls in their own code; they are in the first depth level of the dependency tree.

\begin{table}[H]
	\centering
	\caption{How deep the provider package that raised a manifesting breaking change is from the client in the dependency tree.}
	\begin{tabular}{lrr}
		\toprule
		\textbf{Depth} & \textbf{(\#)} & \textbf{(\%)} \\ \hline
		1              & 27   & 42.2       \\
		2              & 30   & 46.9       \\
		$>$3           & 7    & 10.9
		    \\
		\midrule
		Total & 64 & 100
		\\ \bottomrule
	\end{tabular}
	\label{tab:dependency_tree_deep}
\end{table}

Manifesting breaking changes introduced by indirect providers in the depth level greater than one represent 57.8\% of the cases. Six cases are in the third depth level and a single one is in the fourth depth level. Clients do not install these providers directly; rather, they come from the direct provider. In these cases, the manifesting breaking change may be totally unclear to client packages, since they are typically unaware of such providers (or have no direct control over their installation).

\vspace{5pt}
\begin{mdframed}
\begin{itemize}
    \item The most frequent issues with provider packages that introduced manifesting breaking changes were feature changes, incompatible providers, and object type changes.
    \item Provider packages introduced these manifesting breaking changes at similar rates in minor and patch releases.
    \item Most of the fixed manifesting breaking changes by providers were fixed in patch releases.
    \item Manifesting breaking changes are documented in 78.1\% of the cases, mainly on issue reports.
    \item Indirect providers introduced manifesting breaking changes in most cases.
\end{itemize}
\end{mdframed}


\subsection{RQ3. How do client packages recover from a manifesting breaking change?}

\noindent
\textbf{Finding 8: Clients and transitive providers recover from breaking changes in 39.1\% of cases.}
In the dependency tree, the transitive provider is located between the provider that introduced the manifesting breaking change and the client where it manifested (See Section~\ref{subsec:definitions}). Table \ref{tab:package_fix} shows which package fixed/recovered  from each manifesting breaking change case. The provider packages fixed the majority of the manifesting breaking changes. Since they introduced the breaking change, theoretically this was the expected behavior. Client packages recovered from the manifesting breaking change in 20.3\% of cases, and transitive providers recovered from manifesting breaking changes in 18.8\% of cases. 
When the provider who introduced a manifesting breaking change does not fix it, the transitive provider may fix it and solve the client's issue.

\begin{table}[H]
	\centering
	\caption{Packages fixing/recovering from the error.}
	\begin{tabular}{lrr}
		\toprule
		\textbf{Fixed by/Recovered from}            & \textbf{(\#)} & \textbf{(\%)} \\ \hline
		Provider                     & 32            & 50          \\
		Client                       & 13            & 20.3        \\
		Transitive provider          & 12            & 18.8        \\
		Client + Transitive provider & 25            & 39.1        \\
		Not fixed/recovered                    & 7            & 10.9        \\ 
		\midrule
		Total & 64 & 100 \\
		\bottomrule
	\end{tabular}
	\label{tab:package_fix}
\end{table}

Since transitive providers are also clients of the providers that introduced the manifesting breaking change, clients (clients and transitive providers) recovered from these breaking changes in 39.1\% of cases. This observation suggests that client packages occasionally have to work on a patch when a manifesting breaking change is introduced since in 39.1\% of the cases clients and transitive providers need to take actions to recover from the manifesting breaking change.

\noindent
\textbf{Finding 9: \textit{Transitive providers fix manifesting breaking changes faster than other packages:}} 
When a manifesting breaking change is introduced, it should be fixed by either the provider who introduced it or a transitive provider. In a few cases, the client package will also recover from it. Table \ref{tab:fix_day} shows the time that each package takes to fix the breaking change. In general, manifesting breaking changes are fixed in seven days by provider packages. Even in this relatively short period of time, many direct and indirect clients are affected.

\begin{table}[H]
	\centering
	\caption{Median of number days that each package spent to fix/recover from the manifesting breaking change.}
	\begin{tabular}{lr}
		\toprule
		\textbf{Fixed by/Recovered from}            & \textbf{Days} \\ \hline
		Provider                     & 7             \\
		Client                       & 134           \\
		Transitive provider          & 4             \\ 
		Client + Transitive provider & 82.4 \\ \bottomrule
	\label{tab:fix_day}
	\end{tabular}
\end{table}

Transitive providers fix manifesting breaking changes faster than clients and even providers. Since the manifesting breaking change only exists when it is raised in the client packages, transitive providers break first and need a quick fix; transitive providers usually spent four days to fix a break. Meanwhile, providers that introduced the manifesting breaking change take a median of 7 days to introduce a fix. In cases where the provider neglected to introduce a fix or took longer than the client, client packages took a comparably lengthy 134 days (mean 286; SD 429) to recover from a manifesting breaking change. According to Table \ref{tab:package_fix}, the direct providers and transitive providers fixed most of the manifesting breaking changes, about 78.8\%, because clients can be slow to recover.

However, because transitive providers are also clients, we can analyze the time that clients and transitive providers spend to fix/recover from a manifesting breaking change. Clients and transitive providers recovered from a manifesting breaking change in around 82 days.

\noindent
\textbf{Finding 10: \textit{Upgrading is the most frequent way to recover from a manifesting breaking change.}}
Table \ref{tab:version_change} describes how clients recovered from breaking changes. In 48 cases, the provider version was changed. In most cases (71.4\%), client packages upgraded their providers' version. We analyzed all cases where clients and transitive providers recovered from the manifesting breaking change by changing the provider's version before the provider fixed the error. We observed an upgrade in 12 (52.2\%) cases out of 23. Thus, in more than half of the cases where the client and transitive providers fixed/recovered from the manifesting breaking change, the provider package had newer versions, but the client was not using any follow-up releases from the provider packages.

\begin{table}[H]
	\centering
	\caption{How client packages changed the provider's version after a manifesting breaking change.}
	\begin{tabular}{lcccccccccccc} \toprule
		\textbf{Changed by} & \textbf{Total} & \multicolumn{2}{c}{\textbf{Upgrade}} & \phantom{ab} & \multicolumn{2}{c}{\textbf{Downgrade}} & \phantom{ab} & \multicolumn{2}{c}{\textbf{Replace}} & \phantom{ab} & \multicolumn{2}{c}{\textbf{Remove}}
		\\ \cmidrule{3-4} \cmidrule{6-7} \cmidrule{9-10} \cmidrule{12-13}
		           &    & (\#) & (\%) && (\#) & (\%) && (\#) & (\%) && (\#) & (\%) \\ \midrule
		Client     & 28 & 20   & 71.4 && 6   & 21.4 && 1   & 3.6  &&  1  & 3.6  \\
		Transitive provider & 20 & 9   & 45 && 10   & 50 && 01   & 5  && \textemdash & \textemdash \\ \bottomrule
		\label{tab:version_change}
	\end{tabular}
\end{table}

The number of downgrades in a transitive provider may explain why they recover from the manifesting breaking change faster than the client packages. Since transitive providers are also providers, they should fix the manifesting breaking change as soon as possible, avoiding the propagation of the error caused by the manifesting breaking change. Consequently, the downgrade to a stable release of the provider is the most frequent way for transitive providers to recover from a manifesting breaking change. Finally, the provider is replaced or removed in a small proportion when a breaking change is raised---about 7.2\% for both cases combined. 

\begin{figure}[H]
     \centering
     \begin{subfigure}[b]{0.45\textwidth}
         \centering
         \includegraphics[width=\textwidth]{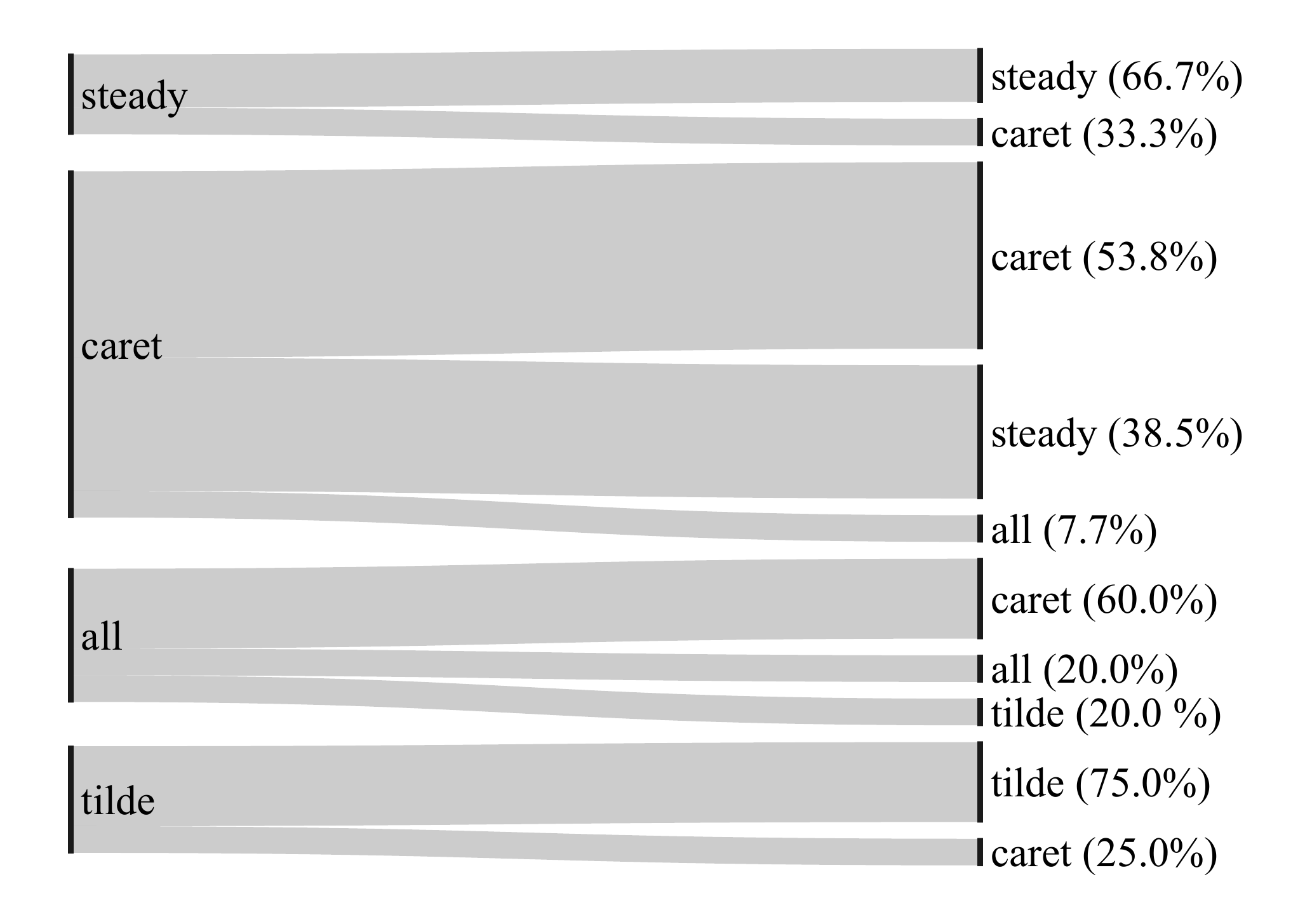}
         \caption{Client}
         \label{fig:semver_both_client}
     \end{subfigure}
     \hfill
     \begin{subfigure}[b]{0.45\textwidth}
         \centering
         \includegraphics[width=\textwidth]{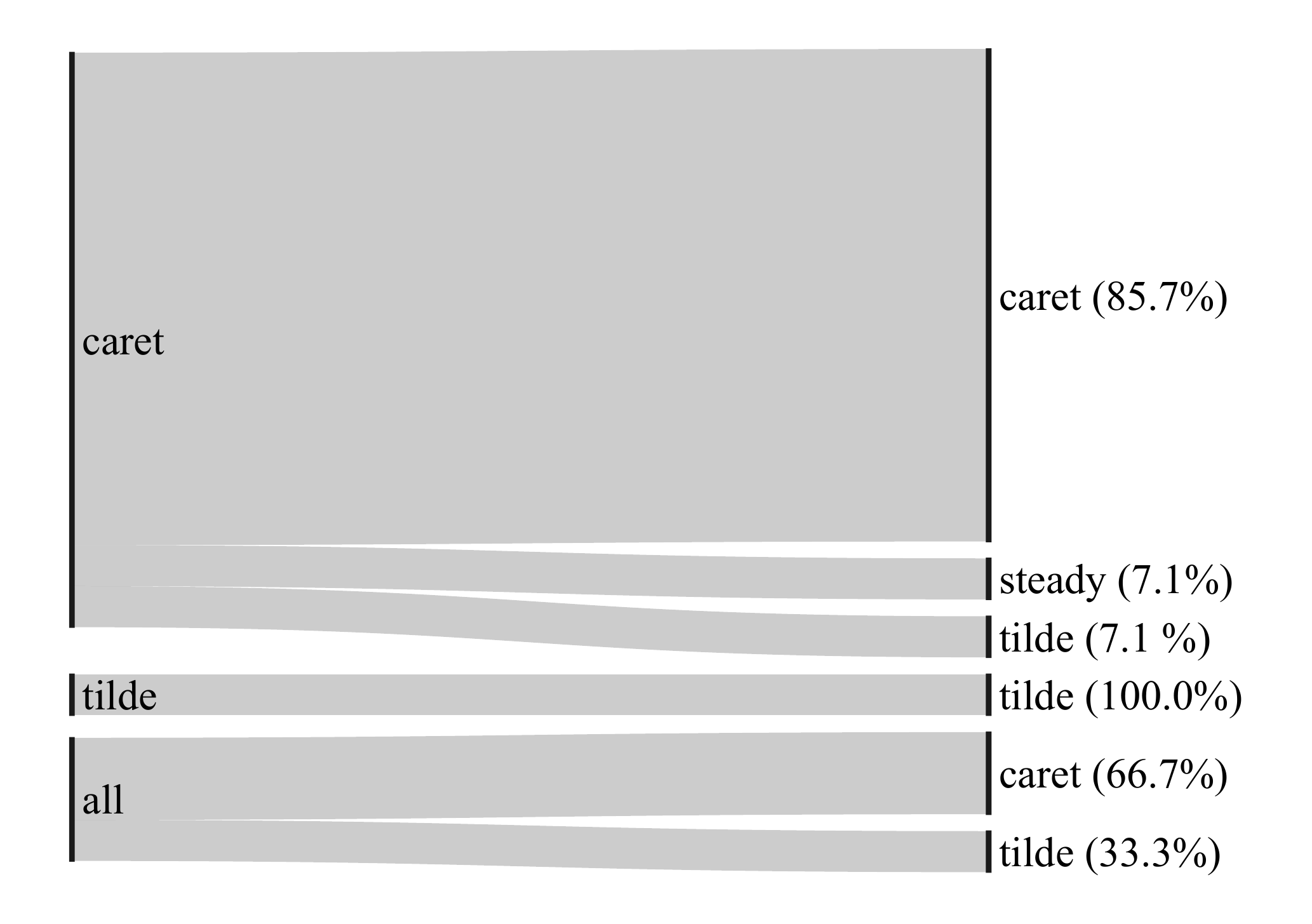}
         \caption{Transitive Provider}
         \label{fig:semver_both_provider}
     \end{subfigure}
    \caption{Provider's version changed by clients and transitive providers. On the left side of each figure, one can see the range level where the manifesting breaking change was introduced and on the right side, one can see the range level where the same manifesting breaking change was fixed.}
    \label{fig:semver_both}
\end{figure}

\noindent
\textbf{Finding 11: \textit{To recover from manifesting breaking changes, clients often change the adopted provider version without changing the range of automatically accepted versions.}}
When a breaking change manifests itself, clients often update the provider's version. Figure \ref{fig:semver_both} shows when the clients and transitive providers updated their providers' versions.

We verified that transitive providers never set a steady version of their provider. When a breaking change manifests in transitive providers, they use a range in the provider's version. However, a single transitive provider changed the range from a caret range to a steady one (e.g., \textasciicircum1.2.1 $\rightarrow$ 1.2.1), to recover from the manifesting breaking change. Nevertheless, when the clients used a caret range and a breaking change manifested, in 38.5\% of the cases they downgraded the provider to a steady version. 

The majority of the manifesting breaking changes were introduced when the clients and transitive providers used the \textit{caret} range (\textasciicircum). It is the default range statement that \textsf{npm} inserts in the \textit{package.json} when a provider is added as a dependency of a client package. In more than half of the cases, these clients changed the provider's version to another caret range. The \textit{accept all} ranges ($>$=, or \textit{*}) were less commonly used and less common when updating.

Clients and the transitive provider in 60.5\% of cases retained the range type and updated it. The range type (all, caret, tilde, or steady) was kept, but the provider was updated/downgraded. For example, a client package specifies a provider \textsf{p@\textasciicircum1.2.0} and receives a breaking change in \textsf{p@1.3.2}. Whenever the provider fixes the code, the client package will update it to, for example, \textsf{p@\textasciicircum1.4.0}, but will not change it for another range type, such as \textit{all}, \textit{tilde}, or \textit{steady} range.

\vspace{5pt}
\begin{mdframed}
\begin{itemize}
    \item Client packages recovered manifesting breaking changes in 39.1\% of cases, including clients and transitive providers.
    \item Providers fixed manifesting breaking changes faster than client packages recovered from manifesting breaking changes by updating the provider, and clients preferred to update rather than downgrade their providers.
    \item The provider's range can be updated or downgraded after a breaking change, but in around 60\% of cases, they did not change the range type.
\end{itemize}
\end{mdframed}


\section{Discussion}

This section discusses the implications of our findings for dependency management practices (Section~\ref{sub:dep_manag}) and the best practices that clients and providers can follow to mitigate the impact caused by manifesting breaking changes (Section~\ref{sub:best_practice}). We also discuss the manifestation of breaking changes and the aspects of Semantic Versioning in the \textsf{npm} ecosystem (Section~\ref{sub:general_discussions}).

\subsection{Dependency management}
\label{sub:dep_manag}
When managing dependencies, client packages can use dependency bots in \textsf{GitHub}, such as \textsf{Snyk} and \textsf{Dependabot}, to receive automatic pull requests when there is a new provider's release~\cite{wessel2018power}. These bots continuously check for new versions and providers' bugs/vulnerabilities fixes. They open pull requests in the client's repository, updating the \textit{package.json}, including changelogs and information about the provider's new version. \citet{automated_pull} show that packages using such bots update their dependencies 1.6x faster than through manual verification. Additionally, tools such as \textsf{JSFIX}~\cite{jsfix2021} can be helpful when upgrading provider releases, especially those that include manifesting breaking changes or major releases. The \textsf{JSFIX} tool was designed to adapt the client code to the new provider release, offering a safe way to upgrade providers.

We verified that a small percentage of the clients recovered from manifesting breaking changes by removing or replacing the provider (c.f., Finding 10), which may be difficult when several features or resources from the provider package are used by the client~\cite{automating_detecting_third-part_java_migration}. Instead, client packages tend to temporarily downgrade to a stable provider version. To ease the process to upgrade/downgrade providers and avoid surprises, clients should search in the provider changelogs for significant changes. As we verified in Finding 6, most manifesting breaking changes are documented in changelogs, issue reports, or pull requests. Dependency bots also could analyze the content of changelogs and issue reports to create red flags, like notifications, about documentation that cites a manifesting breaking change.

Finally, client packages may use a \textit{package-lock.json} file to better manage dependencies. We observed in Finding 7 that indirect providers -- the ones in depth two and three in the dependency tree -- are responsible for 57.8\% of the manifesting breaking changes that affect a client package. Using a \textit{package-lock.json} file, client packages can stay aware of all of the providers' versions of the latest successful build. When a provider is upgraded due to the range of versions and the new release manifests a breaking change on the client side, the client can still install all of the providers' versions that successfully built the client.

\subsection{Best practices}
\label{sub:best_practice}

Several issues found in our manual classification of manifesting breaking changes (Section~\ref{sub:rq2}) could be avoided through the use of static analysis tools. Errors classified as \textit{Semantically Wrong Code} and \textit{Rename function} are typically captured by such tools. Both client and provider developers can use such tools. For a dynamic language such as \textsf{JavaScript}, these tools can help avoid some issues~\cite{js-linters}. Options for \textsf{JavaScript} include \textsf{jslint}, \textsf{jshint} and \textsf{standard}. \citet{js-linters} and \citet{js-linters2} show that developers use linters mainly to prevent errors, bugs, and mistakes.

Due to the dynamic nature of \textsf{JavaScript}, however, static analysis tools cannot verify inherited objects' properties. They do not capture errors classified as \textit{Change one rule}, \textit{Object type change}, and \textit{Undefined object}, as well as \textit{Rename Function} in functions of object's properties. Thus, developers should be concerned about creating test cases that run their code along with the functionality of providers, as only then will they (client developers) find breaking changes that affect their own code. Many available frameworks, such as \textsf{mocha}, \textsf{chai}, and \textsf{ava}, support these tasks. These tests should also be executed on integrated environments every time the developer commits and pushes new changes. For this case, several tools are available, such as \textsf{Travis}, \textsf{Jenkins}, \textsf{Drone CI}, and \textsf{Codefresh}. Using linters and continuous integration systems, developers can catch most of these errors before releasing a new version.

Finally, a good practice for \texttt{npm} packages is to keep a changelog or to document breaking changes and their fixes in issue reports and pull requests. This practice should continue and be more widely adopted, since currently around a fifth of providers do not do it (c.f., Finding 6). This would also help the development of automated tools (e.g. bots) for dealing with breaking changes. Providers could create issue reports and pull request templates to allow clients to specify consistent descriptions of issues they found.
\subsection{Breaking changes manifestation and Semantic Versioning}
\label{sub:general_discussions}

Breaking changes often occur in the \textsf{npm} ecosystem and impact client packages (c.f., Finding 1). Most of the manifesting cases come from indirect providers; that is, providers from the second level or deeper in the dependency tree. Findings from~\citet{npm-three} show that in 2016 half of the client packages in \textsf{npm} had at least 22 transitive dependencies (indirect providers), and a quarter had at least 95 transitive dependencies. In this context, clients may face challenges in diagnosing where the manifesting breaking changes came from, because when a manifesting breaking change is introduced by an indirect provider, the client may not know this provider.

Our results show that provider packages introduce manifesting breaking changes in minor and patch levels, which in principle should only contain backward-compatible updates according to the Semantic Versioning specification. Semantic Versioning is a recommendation that providers can choose to use it or not~\cite{TosemBreaking, what-dependencies-tell-semver}. If providers do not comply with Semantic Versioning, several errors might be introduced, as we observed in Finding 4 that all manifesting breaking changes in pre-releases were propagated to stable releases (c.f., Finding 4). One hypothesis is that providers might be unaware of the correct use of the Semantic Versioning rules, which may explain why they propagated the unstable changes to stable releases. Finally, \textsf{npm} could provide \textit{badges} where provider packages would be able to explicitly show that they are aware of and adhere to the Semantic Versioning. \citet{badges-on-npm-github} claims that developers use visible signals (specifically on GitHub) like badges to indicate project quality. This way, clients could make a better choice about their providers and prefer those aware of Semantic Versioning.


\section{Related work}
\label{sec:related}

This section describes related work regarding breaking changes in \textsf{npm} and other ecosystems.

\noindent
\textbf{Breaking changes in \textsf{npm}:}
\citet{when_it_breaks} presents a survey about the stability of dependencies in the \textsf{npm} and \textsf{CRAN} ecosystem. The authors interviewed seven package maintainers about software changes. In this paper, interviewees highlighted the importance of adhering to Semantic Versioning to avoid issues with dependency updates. More recently, the authors investigated policies and practices in 18 software ecosystems, finding that all ecosystems share values such as stability and compatibility, but differ on other values \cite{TosemBreaking}. \citet{detecting_bc_JavaScript_apis} studied API breaking changes in three provider packages. The author uses \textit{3k} client packages, parsing the providers' and clients' files to detect API-breaking changes and their impact on clients. This work identified that 9.8\% to 25.8\% of client releases are impacted by API-breaking changes.

\citet{noregrets2018} present a technique called \textit{type regression testing} that verifies the type of a returned object from an API and compares it with the returned type in another provider release. The authors chose the 12 most popular provider packages and their major releases, applying the technique in all patch/minor releases belonging to the first major update. They verified type regression in 9.4\% of the minor or patch releases. Our research focused on any kind of manifesting breaking changes and we analyzed both client and provider packages, with 13.9\% of releases impacted by manifesting breaking changes. 

\citet{using_others_tests} focus on detecting break-inducing versions of third-party dependencies. The authors analyzed \textit{290k} \textsf{npm} packages. They flagged each downgrade in the provider version as a possible breaking change. These provider versions were tested using client tests and the authors identified 4.1\% of fails after an update, which resulted in a downgrade. Similar to these authors, we resolved each client's providers for a release, but we ran the tests whenever at least one provider version changed.

\citet{tapir2021} present a tool that uses breaking change patterns described by providers and fixes the client code. They analyzed a dataset with ten of the most used \textsf{npm} packages and searched for breaking changes described in changelogs. We can compare our classification (Finding 3) with theirs. They found 153 cases of breaking changes that were introduced in major releases. They claim that most of the breaking changes (85\%) are related to specific package API points, such as \textit{modules}, \textit{properties}, and \textit{function} changes. Considering our classification (Finding 3), \textit{feature changes}, \textit{object type changed}, \textit{undefined object}, and \textit{renamed function} can also be classified as changes in the package API and, if so, we claim that 64.06\% of manifesting breaking changes are package API related. 

\noindent
\textbf{Breaking changes in other ecosystems:}
\citet{why_how_java} studied 400 providers from the \textsf{Maven} repository for 116 days. The provider packages were chosen by popularity on \textsf{GitHub} and the authors looked for commits that introduced an API-breaking change during that period. Developers were asked about the reasons for breaking changes that occurred. Our paper presents similar results: the authors claim that \textit{New Feature} is the most frequent way a breaking change is introduced, while we claim that Feature Change is the main breaking change type (Finding 3). Also, the authors similarly detected that breaking changes are frequently documented on changelogs (Finding 6).

\citet{efficient_static_checking} present a study about API breaking changes in the \textsf{Maven}, \textsf{PyPI}, and \textsf{RubyGems} ecosystems. The study focuses on detecting breaking changes by computing a \textit{diff} between the code of two releases. They found API-breaking changes in 26\% of provider packages, and their approach suggests automatic upgrades for 10\% of the packages. Our approach goes beyond API breaking changes; we found that 11.7\% of the client packages are impacted by manifesting breaking changes.


\section{Threats to validity}
\label{sec:threats}
\textbf{Internal validity:} 
When a breaking change was detected, we verified the type of change that the provider package introduced and collectively grouped the changes into categories. However, some cases might fall into more than one category. For example, a provider package changes the type of an object to change/improve its behavior. This case might fall into \textit{Feature change} and \textit{Object type changed}. So, we categorized the case in the category that most represents the error. In this case, since the object is changed by a feature change, the most appropriate category would be \textit{Feature change}.

The error cases that we categorized as \textit{breaking due to external change} are the ones in which the clients or providers use -- or depend on -- external data/resources from sites and APIs that changed over time (see Finding 1). These cases represent about 8.1\% of the client's releases, and, in these cases, we could not search for manifesting breaking changes because we could not execute the release tests. After all, the data/resource needed by the test were no longer available. So, about 8\% of client releases might be impacted by breaking changes, but we could not analyze them. 

\noindent
\textbf{Construct validity:} 
In our approach to detecting breaking changes, we only performed an analysis when the client tests failed. If a client used a provider version that had a breaking change, but the client did not call the function that causes the breaking change or did not have tests to exercise that code, we could not detect the breaking change. This is why we call all of our cases \textit{manifesting breaking changes}.

Therefore, we might not have detected all API-breaking changes, as we were able to detect only API name changes and API removal. Parameter changes may not be detected because \textsf{JavaScript} allows making a call to an API with any number of parameters.\footnote{https://eloquentJavaScript.net/03\_functions.html\#p\_kzCivbonMM}

We restored the working tree index in the respective commit tagged by the developer for each release. We listed all tags in the repository, and we used the \textit{checkout} with the respective tag. However, for untagged releases we performed a checkout in the timestamp referenced in the \textit{package.json}. We trusted the timestamp once we verified that the tags and timestamp point to the same commit in 94\% of cases for tagged repositories.

Lastly, we did not mention the file \textsf{npm-shrinkwrap.json} in our study. This file is intended to work like the file \textsf{package-lock.json} when controlling transitive dependency updates, but it may be published along with the package. However, \textsf{npm} strongly recommend avoiding its use. Also, the existence of \textsf{npm-shrinkwrap.json} files does not play any major role in our study, as they do not affect our results, based on our adopted research method. We did not include them in our study.

\noindent
\textbf{External validity:} 
We randomly selected client packages that varied in release numbers, clients, providers, and size. However, since we only analyzed \textsf{npm} packages hosted at GitHub projects, our findings cannot be directly generalized to other settings. It is also important to state that representativeness can also be limited because \textsf{npm} increases the number of packages and releases daily. Future work can replicate our study in other platforms and ecosystems. Finally, since the number of projects in our sample is small, we do not have enough statistical power to perform hypothesis tests around results that involve package-level comparisons.

\noindent
\textbf{Conclusion validity:} 
Conclusion validity relates to the inability to draw statistically significant conclusions due to the lack of a large enough data sample. However, as our research used a qualitative approach, we mitigate any potential conclusion threat by conducting a sanity check on repositories of all client packages with fewer than four releases. This guarantees that all packages are intended for use in production (Subsection \ref{subsubsec:identifying_breaking_changes}). Finally, all of the manifesting breaking changes that we claim in our work were manually analyzed to assure they are legitimate breaking changes that impact clients in the real world (Subsection \ref{subsection:sanity-check-failure-cases}).


\section{Conclusions}
\label{sec:conclusions}

Software reuse is a widely adopted practice, and package ecosystems such as \textsf{npm} support reusing software packages. However, breaking changes are a negative side effect of software reuse. Breaking changes and their impacts are studied in the literature in several software ecosystems \citep{why_how_java, model_based_bc_nodejs, how_to_break_an_api, change_contracts}. A few papers examine breaking changes in the \textsf{npm} ecosystem from the client packages perspective, i.e., executing the client tests to verify the impact of breaking changes \citep{when_it_breaks, noregrets2018, using_others_tests}. In this work, we analyzed manifesting breaking changes in the \textsf{npm} ecosystem from the client and provider perspectives, providing an empirical analysis regarding breaking changes in minor and patch levels.

From the client's perspective, we analyzed the impact of manifesting breaking changes. We found that 11.7\% of clients are impacted by such changes and offer some advice to help clients and automated tools developers discover, avoid, and recover from manifesting breaking changes. Clients can use dependency bots to accelerate the process of upgrading their providers, and clients can look at changelog files for any non-desired updating, such as breaking changes. From the provider's perspective, we analyzed the most frequent causes of manifesting breaking changes. We found that the most common causes were when providers changed some rules/behaviors on features that had been stable over the last releases, when an object type changes, and when there were unintentionally undefined objects at runtime. 
Maintainers should pay attention during code review phases regarding these issues. 
Future research can look into the correlation among package characteristics and metrics with breaking change occurrence.

\section{Acknowledgments}
This work is partially supported by the National Science Foundation under Grant Number IIS-1815503, CNPq/MCTI/FNDCT (grant \#408812/2021-4 ) and MCTIC/CGI/FAPESP (grant \#2021/06662-1).

\bibliographystyle{ACM-Reference-Format}
\bibliography{references}

\end{document}